\documentclass{article}

\usepackage{amsmath}

\usepackage[frozencache]{minted}

\usepackage{fvextra} % Enhances minted's verbatim features

\usepackage{hyperref}
\usepackage{url}
\usepackage{cleveref}

\usepackage{graphicx}
\usepackage{cite}

\usepackage{easyReview}

\setminted{
	linenos,
	breaklines,
	fontsize=\tiny,
	bgcolor=gray!10,
	frame=lines,
	framesep=2mm,
	baselinestretch=1.2,
	tabsize=4,
    obeytabs,        % Respect tabs
    linenos,
    escapeinside=||,
}

\newcommand*{\pegref}[1]{\cref{#1} line \ref{#1}}

\title{
	An Interoperable Syntax for Gas Scattering Reaction Definition
}
\author{
	Dan Andrei Ciubotaru\thanks{These authors contributed equally to this work and are listed alphabetically.}, 
	Michele Renda\footnotemark[1], 
	Călin Alexa\thanks{Corresponding author: \texttt{calin.alexa@cern.ch}}
    }

\date{\today}

\begin{document}

\maketitle 

\begin{abstract}
    We propose a unified, human-readable, machine-processable novel syntax/notation designed to comprehensively describe reactions, molecules and excitation states. Our notation resolves inconsistencies in existing data representations and facilitates seamless integration with computational tools. We define a structured syntax for molecular species, excitation states, and reaction mechanisms, ensuring compatibility with a wide range of scientific applications. We provide a reference implementation based on Parsing Expression Grammar syntax, enabling automated parsing and interpretation of the proposed notation. This work is available as an open-source project, enabling validation and fostering its adoption and further improvement by the scientific community. Our standardized framework provides gas scattering models with increased interoperability and accuracy.
\end{abstract}

\section{Introduction}
In recent years, there has been a renewed interest in gas-based particle detectors due to their sensitivity, fast response time, relatively low cost, and remarkably large sensitive area. This resurgence has led to increased efforts in studying existing detector layouts and designing new ones that offer improved performance and lower costs. However, one of the challenges in designing gaseous detectors is that the detection and amplification mechanisms are intrinsically stochastic and highly sensitive to modifications in the operating setup, such as electric and magnetic fields, geometries, gas mixtures, and impurities.

Designing and benchmarking such detectors using only theoretical models is highly challenging. Consequently, researchers rely on stochastic models to analyze and characterize their behavior. Several software packages have been developed to assist in this work, including \texttt{Magboltz} \cite{magboltz}, \texttt{Bolsig} \cite{bolsig}, \texttt{MCIG} \cite{mcig}, \texttt{Methes} \cite{methes}, and \texttt{LoKI-B} \cite{lokib}/ \texttt{LoKI-MC}\cite{lokimc1, lokimc2}.

A demanding problem for cross-validating software tools is the difficulty of interchanging cross-section tables between different tools, which makes it challenging to decouple the input data (cross-section tables) from the calculation algorithms. Regardless of the approach we use, performing calculations — whether through Monte Carlo or the Boltzmann method — we found that all these tools require access to a set of cross-section tables for the processes of interest.

The LxCAT project \cite{lxcat, lscat_online} has made significant progress in collecting and centralizing cross-section data from various experiments, literature sources, and software tools. 

However, several issues hinder using these tables across different software tools. More specifically:
\begin{itemize}
	\item The reactions are written in a human-readable syntax that software tools cannot automatically process.
	\item There is no standardized syntax for excitation states.
	\item The final state is often defined as an unknown state or a range of states.
\end{itemize}

In an attempt to resolve these issues, we propose a common syntax for reactions, molecules, and excitation states that is both human-readable and unambiguously defined, allowing it to be processed by software tools. 

In \cref{sec:methodology}, we outline the principles used to define the proposed notation in \cref{sec:proposal}. In \cref{sec:specie}, we define the notation for molecules, particles, and species. In \cref{sec:excitations}, we introduce a standardized syntax for describing molecular and atomic excitation states. Finally, in \cref{sec:reactions}, we integrate all defined entities to construct reactions that can be used to describe scattering processes in cross-section tables.
To demonstrate the feasibility and practicality of the proposed notation, in \cref{sec:reference_implementation}, we present a reference implementation that effectively utilizes the Parsing Expression Grammar (PEG) file introduced in \cref{sec:peg} to parse any entity described in this article.

\section{Methodology} \label{sec:methodology}

Defining a notation that is accepted by the scientific community is quite a challenging task. We aimed to follow as closely as possible the notation found in existing literature and software tools whenever feasible. When multiple notations are available (e.g., \texttt{O2\^{}+\^{}+}, \texttt{O2++}, or \texttt{O2\^{}2+}), we adopted a pragmatic approach, choosing to support all commonly used syntaxes.

During the early stages of notation development, we recognized the need for a precise definition of the proposed syntax. Due to the complexity of the task — particularly the need to support ranges of states (see \cref{sec:properties}), we realized that regular expressions alone were insufficient for our purposes. Consequently, we decided to define a fully structured language syntax that could be used to create an effective parser. We opted for a PEG syntax due to its expressiveness and built-in prioritization for resolving ambiguous notations, which ultimately provided an efficient and practical solution.

To manage systematically potential inaccuracies, missing attributes, etc., we have adopted a semantic versioning system using three numbers to uniquely identify versions: \texttt{major.minor.patch}.

\begin{itemize}
	\item A change in the \texttt{patch} number will occur when only minor clarifications or refinements are made to interpret this reference syntax. In this case, interoperability will be fully preserved both forward and backward.
	\item A change in the \texttt{minor} number introduces new syntax elements and/or keywords while maintaining compatibility with the existing ones. Forward compatibility (support for the new syntax by older software implementations) is not guaranteed, but backward compatibility (support for the old syntax by newer software implementations) will still be ensured.
	\item A change in the \texttt{major} number signifies a fundamental revision of the syntax that breaks both backward and forward compatibility.
\end{itemize}

The syntax defined in this article will be designated as \texttt{v1.0.0} and will serve as the basis for the reference implementation described in \cref{sec:reference_implementation}.

\section{Proposal syntax/notation} \label{sec:proposal}

\subsection{Common syntax/notations}

A syntax/notation requires a good balance between readability and rigor of its definition. This article aims to achieve both goals by using commonly accepted expressions to describe each entity while providing direct references to the PEG syntax, where the exact definitions can be found (see \cref{sec:peg}).

Each line of the PEG definitions specifies an entity, such as a molecule or particle. In the article, every time we introduce a new entity, we highlight it in \textbf{bold font}. After presenting the definition in natural language, we provide examples to illustrate the actual syntax used.

In \cref{sec:peg_common}, we define the fundamental primitive types, such as \textbf{integer}, \textbf{float}, and \textbf{fraction}, as well as the Latin and Greek \textbf{letters} used to represent atomic and molecular excitation levels. All other entities in the PEG files build upon these definitions to form more complex structures while maintaining a simple and clean syntax.

\subsection{Properties and ranges} \label{sec:properties}
When defining electron scattering reactions in low-temperature plasma, we often encounter situations where the final molecular state is not well defined, either being unknown or describing a range of states. For this reason, we decided from the outset to integrate support for value ranges into the syntax, enabling seamless parsing of state ranges.

We accomplished this through the definition of \textbf{properties} (see \pegref{peg:INTEGER_PROPERTY} and \ref{peg:FRACTION_PROPERTY}). Each property may represent a specific value (an integer or a fraction), a bounded or unbounded range, or an unknown value, represented by \texttt{*}. This definition is particularly useful in \cref{sec:peg_exc}, where we replace integer and fractional values in molecular and atomic states with integer and fractional properties, allowing the representation of complex excitation state ranges.

\subsection{Specie definition} \label{sec:specie}
\subsubsection{Particles and elements definition}

A \textbf{particle}, as shown in \pegref{peg:particle}, is defined as an indivisible elementary particle and is specified using its common name: \texttt{e}, \texttt{mu}, \texttt{tau} for the electron, muon, and tauon, respectively, and \texttt{e\^{}+}, \texttt{mu\^{}+}, \texttt{tau\^{}+} for their corresponding antiparticles (however, in practice, only electrons are commonly found in low-temperature plasma physics reactions).

Atoms, hereafter referred to as \textbf{elements}, are specified using their standard chemical symbols, such as \texttt{He}, as defined in \pegref{peg:element}. Two special elements, \texttt{D} and \texttt{T}, represent isotopes commonly found in low-temperature plasma physics, namely \textit{deuterium} and \textit{tritium}, respectively.

\subsubsection{Molecule definition}

One or more elements can combine to form complex \textbf{molecules}, using the conventional chemical notation, such as \texttt{CO2} or \texttt{(NH4)2SO4}. Additionally, an arbitrary prefix can be added to a molecule to distinguish between different isomers, enabling differentiation between, for example, \textit{butane} (\texttt{n-C4H10}) and \textit{isobutane} (\texttt{i-C4H10}). The formal definition of this notation is provided in \pegref{peg:molecule}.

\subsubsection{Specie definition}

A \textbf{molecular specie} is defined as a molecule associated with a specific mutable state, such as ionization or excitation. Ions can be represented using commonly known notations such as \texttt{H2O\^{}+} and \texttt{CO2\^{}-}. Multiple ionizations can be expressed using either \texttt{H2O\^{}+\^{}+} or \texttt{H2O\^{}2+}, as well as the commonly encountered forms \texttt{H2O++} and \texttt{CO2--}. The accepted syntax is formally defined in \pegref{peg:molecule_specie} and \pegref{peg:ion_state}.

Molecular excited states are denoted by enclosing the state in parentheses, separated by at least one space, such as \texttt{CO2 (STATE)}, where \texttt{STATE} corresponds to an excitation state as defined in \cref{sec:excitations}.

For consistency, a \textbf{particle specie} object is also defined, which includes the elementary particle type and quantity, allowing for a compact notation such as \texttt{2e} to represent two electrons produced in a reaction. The formal definition is provided in \pegref{peg:particle_specie}.

\subsection{Excitations} \label{sec:excitations}

In a reaction, each molecule can be found in a state that is different from the ground state due to inelastic interaction with other molecules or particles. Sometimes, the exact state of the molecule is unknown, except that it is in an excited state. In such cases, the literature conventionally denotes the state of the molecule with a \texttt{*} to indicate an undefined excited state, a syntax that we have chosen to preserve: \texttt{H2 (*)}.  

Suppose we decide to dive into the details of the excitation. In that case, we can see that three main mechanisms can alter the internal state of the molecule: \texttt{electronic excitation}, caused by electrons transitioning to higher energy levels; \texttt{vibrational excitation}, resulting from internal movements of molecular components that modify the molecule's geometry, and \texttt{rotational excitation}, which arises from rotational motion that does not alter the molecule's internal geometry.  

Since all these excitation modes can coexist, it is possible to specify each excitation state separated by a comma, provided that they are listed in the correct order: electronic, vibrational, and rotational. If a particular excitation is not present, it can be omitted. For example:  \\
\texttt{
	CO2 (2V1)			\\
	CO2 (1PI,2V1)		\\
	CO2 (2V1,J=1)		\\
	CO2 (1PI,2V1,J=1)	\\	
}
(for details, please see \pegref{peg:exc_state}.)

\subsubsection{Electronic}

The description of electronic excitation is the most complex aspect of this proposed syntax due to the intrinsic complexity of this class of processes and the existence of different coupling schemes used to represent such states. While defining this syntax, it became clear that a single scheme would not be sufficient to describe all possible electronic excitation states found in the literature. For this reason, it was decided not to enforce a single scheme but to promote and properly define the most widely used coupling schemes found in the literature (see \cref{tab:coupling_schemes} for a list of the supported schemes).

\begin{table}[]
	\centering
	\begin{tabular}{cl}
		Code & Description               \\\hline
		\texttt{LS}   & Russell-Saunders coupling \\
		\texttt{JJ}   & jj-coupling               \\
		\texttt{JL}   & jl-coupling               \\
		\texttt{RH}   & Racah coupling            \\
		\texttt{PS}   & Paschen coupling          \\
		\texttt{UU}   & Unspecified excitation   
	\end{tabular}
	\caption{Coupling schemes used to describe electronic excitations}
	\label{tab:coupling_schemes}
\end{table}

\textbf{Russell-Saunders coupling (\texttt{LS})}
This scheme is the most commonly used when spin-spin interactions dominate over orbital-orbital interactions, which, in turn, are stronger than spin-orbit interactions. This condition generally holds for light atoms ($Z \leq 30$) in the absence of strong magnetic fields.  

This notation can be used for both atoms and light molecules, and it is described in \pegref{peg:exc_ele_ls}. An example of this syntax is: \\
\texttt{
	\textit{Single atoms:} \\
	C (3P0)	\\
	C (1D2) \\
	C (1S0) \\
	\textit{Molecules:} \\
	O2 (X 3SIGg-) \\
	O2 (a 1SIGg) \\
	O2 (A 3SIG+) \\
}

\textbf{jj-coupling (\texttt{JJ})}

This scheme is used when the spin-orbit interaction dominates over other interactions, usually in heavy atoms ($Z > 30$). The syntax of this coupling scheme is described in \pegref{peg:exc_ele_jj}. An example of such coupling is: \\
\texttt{
	Pb  ((6p1/2,6p3/2)1) \\
	PbF ((6p3/2,2p3/2)1) \\
}

\textbf{jl-coupling (\texttt{JL})}

The jl-coupling notation is often preferred over LS notation for heavy atoms ($Z > 30$). This notation is described in \pegref{peg:exc_ele_jl}. An example of this scheme is: \\
\texttt{
	Pb  (6p[3/2]3/2) \\
	Tl  (6s[1/2]1/2) \\
}

\textbf{Racah coupling (\texttt{RH})}

Racah notation is used when spin-orbit interactions are significant. This notation is described in \pegref{peg:exc_ele_rh}. It consists of a canonical form, which includes the parent state from which the excitation arises, and a compact form. An example of this scheme in the canonical form is:  \\
\texttt{
	Pb ((3P) 6p[3/2]1) \\
	Pb ((1S) 6p[1/2]1/2) \\
}

However, a compact form also exists, omitting the explicit parent state. In this form, a single apostrophe denotes \texttt{2P3/2}, while its absence indicates \texttt{2P1/2}. Examples include:  \\
\texttt{
	Ne (4s[3/2]1) \\
	Ne (4s'[1/2]0) \\
	Ne (4s[3/2]2) \\
	Ne (4s[3/2]1) \\
}

\textbf{Paschen coupling (\texttt{PS})}

This notation is described in \pegref{peg:exc_ele_ps}. Examples of this notation are:  \\
\texttt{
	He (3s2) \\
	He (2p10) \\
	He (3s6) \\
	He (2p4) \\
}

\subsubsection{Vibrational}

\textbf{Vibration modes}

\begin{figure}
	\centering
	\includegraphics[width=\linewidth]{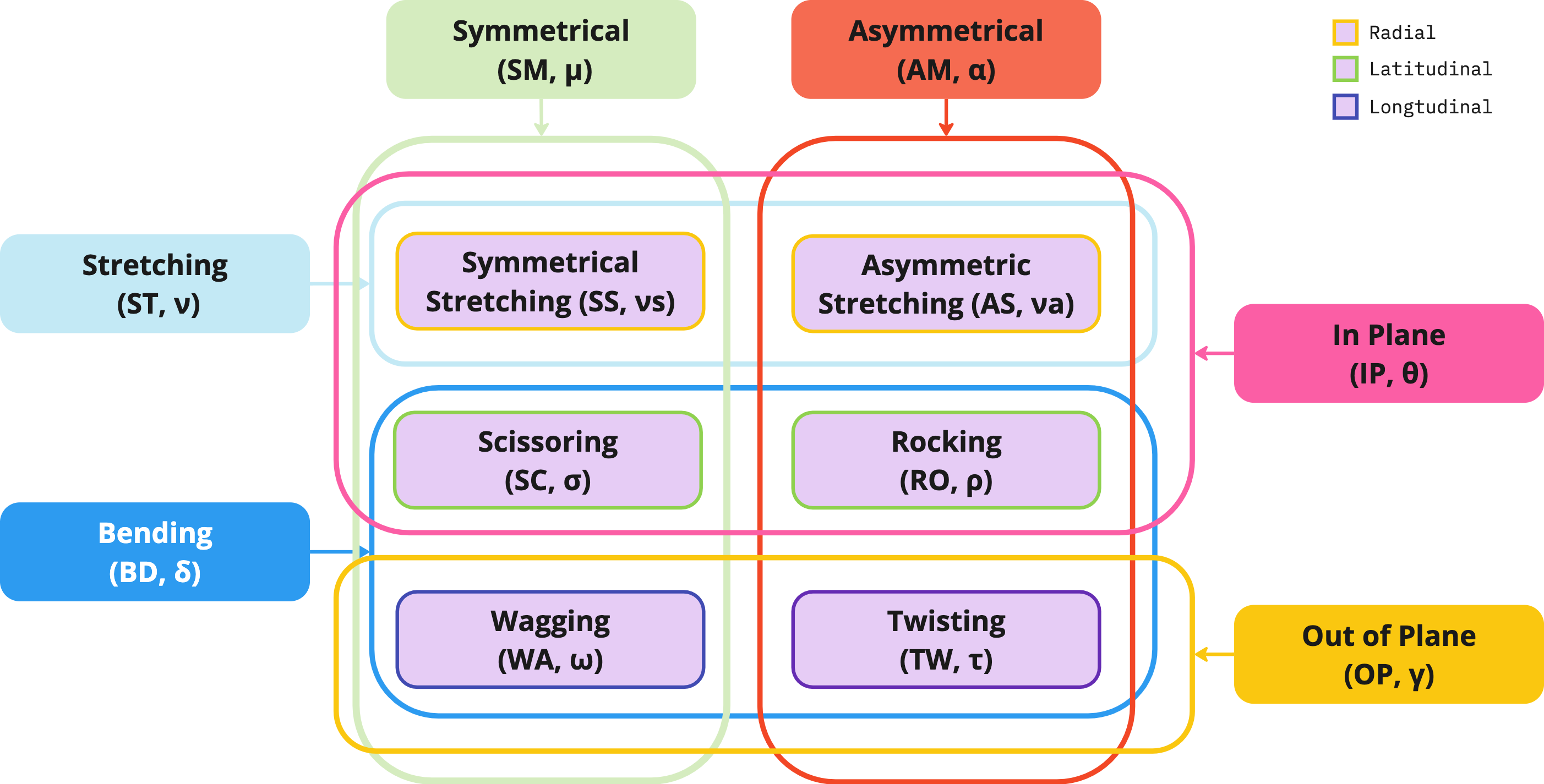}
	\caption{Classification of the vibrational modes: the yellow-highlighted Greek letters represent new symbols introduced in this article, while the non-highlighted ones are already used in the literature, though not always consistently.}
	\label{fig:vibrational_modes}
\end{figure}

For a detailed description of vibrational excitation, the intrinsic vibrational modes of the molecule must be known, as they depend on the geometric structure and the strength of interatomic bonds. These modes are well studied and documented, and our work is limited to compiling and categorizing them, assigning each mode a two-letter code (see \cref{tab:vibration_modes}). Some modes, such as \texttt{ST}, encompass other modes, as illustrated in \cref{fig:vibrational_modes}, and can be used when a more detailed description is unavailable.

Additionally, in the literature, vibrational modes are often represented by a single Greek letter; however, there is no universally defined convention for their use, leading to confusion and potential misunderstandings. In \cref{tab:vibration_modes}, we also specify the corresponding Greek symbol for each mode to eliminate any ambiguity. Nevertheless, our proposed syntax relies exclusively on two-letter Latin codes.

\begin{table}
	\centering
	\begin{tabular}{c c ll}
		Notation 	& Greek Letter & Description 				& Includes \\\hline
		\texttt{SM} & $\mu$		& symmetrical 				& \texttt{SS} \texttt{SC} \texttt{WA}\\
		\texttt{AM} & $\alpha$ 	& asymmetrical 				& \texttt{AS} \texttt{RO} \texttt{TW} \\
		\texttt{SS} & $\nu_s$	& symmetrical stretching 	&  \\
		\texttt{AS} & $\nu_a$	& asymmetrical stretching 	&  \\
		\texttt{ST} & $\nu$		& stretching 				& \texttt{SS} \texttt{AS} \\
		\texttt{BD} & $\delta$ 	& bending 					& \texttt{SC} \texttt{RO} \texttt{WA} \texttt{TW} \\
		\texttt{IP} & $\theta$ 	& in-plane 					& \texttt{SS} \texttt{AS} \texttt{SC} \texttt{RO} \\
		\texttt{OP} & $\gamma$	& out-of-plane 				& \texttt{WA} \texttt{TW}\\
		\texttt{SC} & $\sigma$	& scissoring 				& \\
		\texttt{RO} & $\rho$ 	& rocking 					& \\
		\texttt{WA} & $\omega$	& wagging 					& \\
		\texttt{TW} & $\tau$	& twisting 					& \\
	\end{tabular}
	\caption{Notation for vibration modes}
	\label{tab:vibration_modes}
\end{table}

\textbf{Syntax}

Vibrational excitations are described in \pegref{peg:exc_vib}. An undefined vibrational excitation can be written as \texttt{V*}, a notation congruent with the other notations of undefined excitations.

Due to its nature, multiple vibrational excitations can coexist in the same molecule. These vibrational excitations may interact, forming bands known as Fermi resonances and combination bands.

One or multiple (both interacting or non-interacting) vibrational excitations form a \textbf{excitation vibrational set} (see \pegref{peg:exc_vib_set}). If the excitation is non-interacting, each \textbf{excitation vibration level} is separated by a space (see \pegref{peg:exc_vib_level}), otherwise a \texttt{+} or \texttt{-} sign can be used to mark interacting bands.

Each vibration level composing a band is composed of these attributes in order:
\begin{itemize}
	\item Overtone number, by default 1
	\item Vibration model, such as harmonic (\texttt{N}), polyad (\texttt{P}), by default unspecified
	\item The \texttt{V} character to specify it is a vibration excitation
	\item Vibration number
	\item Vibration mode, as defined in \cref{tab:vibration_modes}, enclosed by \texttt{[ ]}, by default unspecified.
\end{itemize}

Some examples of vibration excitation are: \\
\texttt{
	H2O (V1) \\
	H2O (2V1) \\
	H2O (V2[BD]) \\
	H2O (V1[SS] V2[BD]) \\
	\\
	CO2 (V1[SS]+2V2[BD]) \\
	H2O (V2[BD]+V3[ST]) \\
	CH4 (V3[AS]-V4[BD])
}

Vibrational excitations are described in \pegref{peg:exc_vib}. An undefined vibrational excitation can be written as \texttt{V*}, a notation congruent with the other notations of undefined excitations.

Due to their nature, multiple vibrational excitations can coexist in the same molecule. These vibrational excitations may interact, forming bands known as Fermi resonances and combination bands.

One or multiple (both interacting or non-interacting) vibrational excitations form an \textbf{excitation vibrational set} (see \pegref{peg:exc_vib_set}). If the excitations are non-interacting, each \textbf{excitation vibration level} is separated by a space (see \pegref{peg:exc_vib_level}); otherwise, a \texttt{+} or \texttt{-} sign can be used to mark interacting bands.

Each vibration level composing a band consists of the following attributes in order:
\begin{itemize}
	\item Overtone number (default: 1)
	\item Vibration model, such as harmonic (\texttt{N}) or polyad (\texttt{P}) (default: unspecified)
	\item The \texttt{V} character, specifying it as a vibrational excitation
	\item Vibration number
	\item Vibration mode, as defined in \cref{tab:vibration_modes}, enclosed in \texttt{[ ]} (default: unspecified)
\end{itemize}

In addition, we provide support for a compact notation used to describe the vibrational states of linear molecules (e.g., CO2) and is available when there is no overtone, and the vibration number is less than ten (see \pegref{peg:exc_vib_ijk}): \texttt{J(121)}. Using this notation, the first digit represents symmetrical stretching, \texttt{SS}, the second one the bending, \texttt{BD}, while the last number represents asymmetrical stretching, \texttt{AS}.

Some examples of vibrational excitations are:\\
\texttt{
	H2O (V1) \\
	H2O (2V1) \\
	H2O (V2[BD]) \\
	H2O (V1[SS] V2[BD]) \\
	\\
	CO2 (V1[SS]+2V2[BD]) \\
	H2O (V2[BD]+V3[ST]) \\
	CH4 (V3[AS]-V4[BD]) \\
	\\
	V(121)
}

\subsubsection{Rotational}

Rotational excitations are described in \pegref{peg:exc_rot}. An undefined rotational excitation can be simply written as \texttt{J*}, a notation commonly found in the literature. For a more detailed description of rotational excitation, the geometry of the molecule must be known in order to identify the rotational axes and their degeneracies. A single-atom molecule, such as \texttt{Ne}, behaves as a point-like particle in rotational terms and, therefore, cannot have any rotationally excited states.

However, consider a linear molecule such as \texttt{CO}. We find that it has two independent rotational axes, both perpendicular to the molecular axis, which exhibit inertia and can be in an excited state. The third axis, which is aligned with the direction of the linear molecule, does not exhibit inertia or contribute to rotational excitation. In general, for linear molecules, the two perpendicular axes have the same moment of inertia, making them degenerate. As a result, the rotational state can be described using a single quantum number, commonly expressed using the notation \texttt{J=1}.

If we consider a molecule such as ammonia, \texttt{NH3}, which has a trigonal pyramidal structure, we find that two of its three principal moments of inertia are degenerate. Consequently, describing its rotational energy state requires only two quantum numbers. In the literature, these quantum numbers are the principal rotational quantum number, \texttt{J}, and the projection of angular momentum onto the main axis, denoted as \texttt{K}. In our proposed syntax, this rotational excitation can be expressed using the notation: \texttt{J=2,K=1}.

If a molecule has three distinct non-degenerate and non-zero moments of inertia, then three quantum numbers are needed to describe the rotational state precisely:  
The principal rotational quantum number, \texttt{J}, the projections onto the principal axis, respectively \texttt{Ka} and \texttt{Kc}. This rotational excitation can be expressed using the notation: \texttt{J=3,Ka=2,Kc=1}.

A compact notation is available when all quantum numbers are less than ten, allowing a more concise representation: \\
\texttt{
	J(1)    \\
	J(21)   \\
	J(321)  
}

\subsection{Reactions} \label{sec:reactions}

Once the species are defined, it is possible to construct \textbf{reactions}. While reactions are well known in chemistry, special attention must be given to defining all possible processes occurring in low-temperature plasma, including cases where the reaction outcome is not uniquely determined and depends on probabilistic mechanisms.

For this reason, we first define an \textbf{expression} as the concatenation of different species (molecular or particle species), such as:
\texttt{2e + H + O2} (for the precise definition, see \pegref{peg:expression}).

An expression with an associated probability is called a \textbf{reaction branch}. The probability is enclosed in square brackets and can be represented as a floating-point number, a rational fraction, or a percentage, as defined in \pegref{peg:reaction_branch}.

Multiple reaction branches, separated by the \texttt{or} keyword, define a \textbf{reaction step}, as specified in \pegref{peg:reaction_step}:\\
\texttt{e + CO2 [75\%] or 2e + CO2\^{}+ [20\%] or 3e + CO2\^{}2+ [5\%]}.

Once the reaction steps are established, it becomes possible to fully define \textbf{reactions} by combining different reaction steps using the \texttt{=>}, \texttt{<=}, and \texttt{<=>} symbols to denote forward, backward, and bidirectional reactions, respectively:\\
\texttt{e + H2 => e + H2 (*) => 2e + H2\^{}+ [2/3] or 3e + H2\^{}++ [1/3]}\\
as specified in \pegref{peg:reaction}.

\section{Reference Implementation} \label{sec:reference_implementation}

A new syntax has little value if not implemented to verify consistency and identify potential issues. From the beginning of this syntax's development, it was clear that we should provide a reference implementation to demonstrate that what we describe here can have practical applications.

We have decided to publish the PEG grammar presented in this article at the following address:
\begin{center}
	\url{https://gitlab.com/micrenda/zmoles-peg}
\end{center}
and our reference implementation at:
\begin{center}
	\url{https://gitlab.com/micrenda/zmoles}
\end{center}

The main reason for maintaining these as separate projects is to highlight the independence between our syntax and its implementation, encouraging alternative implementations in different programming languages.

For our reference implementation, we have chosen to build an application using modern \texttt{C++-20}, which functions both as a library and as a command-line utility for fast syntax validation.

The library used to parse PEG files is \texttt{cpp-peglib} \cite{cpppeg}, for which the authors of this article are grateful, as it provides an elegant, fast, and robust implementation.

\section{PEG definition} \label{sec:peg}

\subsection{Common definitions} \label{sec:peg_common}
\begin{minted}{bnf}
# Atomic and molecular letters

ORDINAL_UPPER_LETTER	<- 'X' / 'A' / 'C' / 'D' / 'E' / 'F' / 'G' / 'H' / 'I' / 'J' / 'K' / 'L' / 'M' / 'N' / 'O' / 'P'  |\label{peg:ORDINAL_UPPER_LETTER}|
ORDINAL_LOWER_LETTER	<- 'x' / 'a' / 'c' / 'd' / 'e' / 'f' / 'g' / 'h' / 'i' / 'k' / 'l' / 'm' / 'n' / 'o' / 'p'  |\label{peg:ORDINAL_LOWER_LETTER}|

ATOMIC_UPPER_LETTER  	<- 'S' / 'P' / 'D' / 'F' / 'G' / 'H' / 'I' |\label{peg:ATOMIC_UPPER_LETTER}|
ATOMIC_LOWER_LETTER  	<- 's' / 'p' / 'd' / 'f' / 'g' / 'h' / 'i'  |\label{peg:ATOMIC_LOWER_LETTER}|

GREEK_LETTER		  	<- 'SIG' / 'PI' / 'DEL' / 'PHI' / 'GAM' / 'ETA' / 'IOT' |\label{peg:GREEK_LETTER}|

OPTIONAL_EXCLAMATIONS	<- '!'* |\label{peg:OPTIONAL_EXCLAMATIONS}|

OPTIONAL_SINGLE_QUOTE	<- '\''? |\label{peg:OPTIONAL_SINGLE_QUOTE}|

# Property definition
INTEGER_PROPERTY	<-  ANY / INTEGER_RANGE  / INTEGER |\label{peg:INTEGER_PROPERTY}|
INTEGER_RANGE	    <- '[' INTEGER '-' (INTEGER / HIGH) ']' |\label{peg:INTEGER_RANGE}|
FRACTION_PROPERTY	<-  ANY / FRACTION_RANGE / FRACTION |\label{peg:FRACTION_PROPERTY}|
FRACTION_RANGE	    <- '[' FRACTION '-' (FRACTION / HIGH) ']' |\label{peg:FRACTION_RANGE}|

# Common Keywords
FLOAT				<- [+-]? DIGIT+ '.' DIGIT* |\label{peg:FLOAT}|
INTEGER				<- [+-]? DIGIT+ |\label{peg:INTEGER}|
FRACTION			<- [+-]? DIGIT+ ( '/' DIGIT+ )? |\label{peg:FRACTION}|

HIGH				<- 'H' |\label{peg:HIGH}|
ANY					<- '*' |\label{peg:ANY}|
~COMMA				<- ','
PLUS				<- '+' |\label{peg:PLUS}|
MINUS				<- '-' |\label{peg:MINUS}|

UPPERCASE           <- [A-Z] |\label{peg:UPPERCASE}|
LOWERCASE           <- [a-z] |\label{peg:LOWERCASE}|
DIGIT               <- [0-9] |\label{peg:DIGIT}|

~_    				<- [ \t]+
~END                <-  EOL / EOF
~EOL                <- '\r\n' / '\n' / '\r'
~EOF                <- !.
\end{minted}
\subsection{Specie definitions} \label{sec:peg_specie}
\begin{minted}{bnf}
# Specie
specie				<- molecule_specie / particle_specie |\label{peg:specie}|
molecule_specie		<- molecule ion_state? (_ '(' exc_state ')')? |\label{peg:molecule_specie}|
particle_specie		<- INTEGER? particle |\label{peg:particle_specie}|

# Ionization state
ion_state			<- ion_state_p1 / ion_state_m1 / ion_state_p2 / ion_state_m2 / ion_state_p3/ ion_state_m3 |\label{peg:ion_state}|

ion_state_p1		<- '^'? PLUS{2,} |\label{peg:ion_state_p1}|
ion_state_p2		<- ('^' PLUS)+ |\label{peg:ion_state_p2}|
ion_state_p3		<- '^' INTEGER PLUS |\label{peg:ion_state_p3}|

ion_state_m1		<- '^'? MINUS{2,} |\label{peg:ion_state_m1}|
ion_state_m2		<- ('^' MINUS)+ |\label{peg:ion_state_m2}|
ion_state_m3		<- '^' INTEGER MINUS |\label{peg:ion_state_m3}|

# Molecule
molecule			<- (isomer '-')? molecule_comp+ |\label{peg:molecule}|
molecule_comp		<- (element / ('(' molecule ')')) molecule_comp_qty |\label{peg:molecule_comp}|
molecule_comp_qty	<- INTEGER? |\label{peg:molecule_comp_qty}|
isomer				<- [a-z0-9]+ |\label{peg:isomer}|
element             <- UPPERCASE LOWERCASE{0,2} |\label{peg:element}|

# Particle
particle			<- 'tau^+' / 'tau' / 'mu^+' / 'mu' / 'e^+' / 'e' |\label{peg:particle}|
\end{minted}
\subsection{Excitation definitions} \label{sec:peg_exc}
\begin{minted}{bnf}
exc_state            	<- exc_state_1 / exc_state_2 / exc_state_3 / exc_state_4 / exc_state_5 / exc_state_6 / exc_state_7 |\label{peg:exc_state}|

exc_state_1             <- exc_ele COMMA _ exc_vib COMMA _ exc_rot |\label{peg:exc_state_1}|

exc_state_2             <- exc_ele COMMA _ exc_vib |\label{peg:exc_state_2}|
exc_state_3             <- exc_ele COMMA _ exc_rot |\label{peg:exc_state_3}|
exc_state_4             <- exc_vib COMMA _ exc_rot |\label{peg:exc_state_4}|

exc_state_5             <- exc_ele |\label{peg:exc_state_5}|
exc_state_6             <- exc_vib |\label{peg:exc_state_6}|
exc_state_7             <- exc_rot |\label{peg:exc_state_7}|
\end{minted}
\subsubsection{Rotational} \label{sec:peg_exc_rot}
\begin{minted}{bnf}
# Rotational excitation
exc_rot			<- exc_rot_3 / exc_rot_2 / exc_rot_1 / exc_rot_c3 / exc_rot_c2 / exc_rot_c1 / exc_rot_any |\label{peg:exc_rot}|

exc_rot_1       <- 'J=' INTEGER_PROPERTY  |\label{peg:exc_rot_1}|
exc_rot_2       <- 'J=' INTEGER_PROPERTY _ 'K='  INTEGER_PROPERTY  |\label{peg:exc_rot_2}|
exc_rot_3       <- 'J=' INTEGER_PROPERTY _ 'Ka=' INTEGER_PROPERTY _ 'Kc=' INTEGER_PROPERTY  |\label{peg:exc_rot_3}|

exc_rot_c1		<- 'J' '(' DIGIT ')' |\label{peg:exc_rot_c1}|
exc_rot_c2		<- 'J' '(' DIGIT DIGIT ')' |\label{peg:exc_rot_c2}|
exc_rot_c3		<- 'J' '(' DIGIT DIGIT DIGIT ')' |\label{peg:exc_rot_c3}|

exc_rot_any		<- 'J' ANY  |\label{peg:exc_rot_any}|
\end{minted}
\subsubsection{Vibrational} \label{sec:peg_exc_vib}
\begin{minted}{bnf}
# Vibrational excitation
exc_vib        		<- exc_vib_ijk / exc_vib_set / exc_vib_any |\label{peg:exc_vib}|

exc_vib_set         <- exc_vib_level ( exc_vib_op exc_vib_level)* |\label{peg:exc_vib_set}|
exc_vib_level       <- exc_vib_overtone exc_vib_model 'V' INTEGER_PROPERTY ( '[' exc_vib_mode ']' )? |\label{peg:exc_vib_level}|

exc_vib_overtone	<- INTEGER_PROPERTY? |\label{peg:exc_vib_overtone}|
exc_vib_model		<- 'N' / 'P' / '' |\label{peg:exc_vib_model}|
exc_vib_mode		<- 'ST' / 'BD' / 'SM' / 'AM' / 'IP' / 'OP' / 'SS' / 'AS' / 'SC' / 'RO' / 'WA' / 'TW' |\label{peg:exc_vib_mode}|
exc_vib_op   		<- _ / PLUS / MINUS |\label{peg:exc_vib_op}|

exc_vib_ijk			<- 'V(' DIGIT DIGIT DIGIT ')' |\label{peg:exc_vib_ijk}|
exc_vib_any			<- 'V' ANY |\label{peg:exc_vib_any}|
\end{minted}
\subsubsection{Electronic} \label{sec:peg_ele}
\begin{minted}{bnf}
# Rotational excitation
exc_ele				  	<- exc_ele_bg / exc_ele_ps / exc_ele_rh / exc_ele_jl / exc_ele_jj / exc_ele_any |\label{peg:exc_ele}|
						
exc_ele_bg			  	<- INTEGER_PROPERTY (ATOMIC_LOWER_LETTER / ATOMIC_UPPER_LETTER) INTEGER_PROPERTY OPTIONAL_EXCLAMATIONS _ 'J' '=' INTEGER_PROPERTY |\label{peg:exc_ele_bg}|
exc_ele_ps			  	<- INTEGER_PROPERTY ATOMIC_LOWER_LETTER INTEGER_PROPERTY |\label{peg:exc_ele_ps}|
exc_ele_rh			  	<- exc_ele_rh_full / exc_ele_rh_compact |\label{peg:exc_ele_rh}|
exc_ele_jl			  	<- INTEGER_PROPERTY ATOMIC_LOWER_LETTER '[' FRACTION_PROPERTY ']' INTEGER_PROPERTY  |\label{peg:exc_ele_jl}|
exc_ele_jj			  	<- '(' FRACTION_PROPERTY (COMMA FRACTION_PROPERTY)+ ')' INTEGER_PROPERTY exc_ele_inv_parity |\label{peg:exc_ele_jj}|
exc_ele_ls			  	<- INTEGER_PROPERTY (ATOMIC_LOWER_LETTER / ATOMIC_UPPER_LETTER)  FRACTION_PROPERTY  exc_ele_inv_parity |\label{peg:exc_ele_ls}|
						
exc_ele_rh_full		  	<- '(' INTEGER_PROPERTY (GREEK_LETTER / ATOMIC_UPPER_LETTER) FRACTION_PROPERTY ')' _ INTEGER_PROPERTY (ATOMIC_LOWER_LETTER / ATOMIC_UPPER_LETTER) '[' FRACTION_PROPERTY ']' INTEGER_PROPERTY exc_ele_inv_parity |\label{peg:exc_ele_rh_full}|
exc_ele_rh_compact	  	<- INTEGER_PROPERTY ATOMIC_UPPER_LETTER OPTIONAL_SINGLE_QUOTE '[' FRACTION_PROPERTY ']' exc_ele_inv_parity |\label{peg:exc_ele_rh_compact}|
						
exc_ele_inv_parity	  	<- ('u' / 'g')? |\label{peg:exc_ele_inv_parity}|
exc_ele_ref_symmetry  	<- (PLUS / MINUS)? |\label{peg:exc_ele_ref_symmetry}|

exc_ele_any				<- 'E' ANY |\label{peg:exc_ele_any}|
\end{minted}
\subsection{Reaction definitions} \label{sec:peg_rea}
\begin{minted}{bnf}
reaction				<- reaction_step ( _ direction _ reaction_step )+ |\label{peg:reaction}|
reaction_step			<- reaction_branch ( _ 'or' _  reaction_branch)* |\label{peg:reaction_step}|
reaction_branch  		<- expression ( _ '['  branch_ratio ']' )?  |\label{peg:reaction_branch}|

expression				<- specie ( _ ~PLUS _ specie )+ |\label{peg:expression}|
direction				<- '<->' / '->' / '<-'  |\label{peg:direction}|

branch_ratio			<- branch_ratio_perc / branch_ratio_float / branch_ratio_frac  |\label{peg:branch_ratio}|
branch_ratio_percent	<- (FLOAT / INTEGER) '%' |\label{peg:branch_ratio_percent}|
branch_ratio_float		<- FLOAT |\label{peg:branch_ratio_float}|
branch_ratio_fracion	<- FRACTION |\label{peg:branch_ratio_fracion}|
\end{minted}

\section{Conclusions}
This work proposes a unified, human-readable, and machine-processable novel syntax designed to comprehensively describe reactions, molecules and excitation states. 

We resolve inconsistencies in existing data representations and facilitate seamless integration with computational tools. 

Molecular species, excitation states and reaction mechanisms have received a well-defined structured syntax that ensures compatibility with a wide range of scientific applications. 

Based on Parsing Expression Grammar syntax, our open-source project enables automated parsing and interpretation of the proposed notation. 
Its validation, adoption, and further improvement by the scientific community are entirely accessible.

An essential asset of our standardized framework is establishing increased interoperability accuracy for gas scattering models. 

We believe that the promising results presented in this paper have a significant scientific impact on the scientific community interested in modeling gas scattering processes. Furthermore, this work will likely benefit from further improvements. 

\section*{Acknowledgments}
This work was supported by IFIN-HH under Contract No.~PN 23210104 with the Romanian Ministry of Education and Research.

\bibliography{references}
\bibliographystyle{plain}

\end{document}